\documentclass{osa-article}
\journal{oe}

\usepackage{bm}
\newcommand{\Ai}{\operatorname{Ai}}
\newcommand{\sgn}{\operatorname{sgn}}

\newcommand{\sig}{\operatorname{sig}}
\newcommand{\dif}{\mathrm{d}}

\def\mydel#1{\ignorespaces}

\begin{document}
\title{Independent amplitude and trajectory/beam-width control of nonparaxial %accelerating 
beams}

\author{Michael Goutsoulas,% \authormark{1} 
Raluca-Sorina Penciu, % \authormark{1} 
and Nikolaos K. Efremidis\authormark{*}
}

\address{
%\authormark{1}
Department of Mathematics and Applied Mathematics, University of Crete, 70013 Heraklion, Crete, Greece
% \\
%\authormark{2}Institute of Applied and Computational Mathematics, Foundation for Research and Technology - Hellas (FORTH), 70013 Heraklion, Crete, Greece.\\
}

\email{\authormark{*}nefrem@uoc.gr} %% email address is required
\begin{abstract}
We show that it is possible to generate non-paraxial optical beams with pre-engineered trajectories and designed maximum amplitude along these trajectories. The independent control of these two degrees of freedom is made possible by engineering both the amplitude and the phase of the optical wave on input plane. Furthermore, we come to the elegant conclusion that the beam width depends solely on the local curvature of the trajectory. Thus, we can generate beams with pre-defined amplitude and beam-width by appropriately selecting the local curvature. Our theoretical results are in excellent agreement with numerical simulations. We discuss about methods that can be utilized to experimentally generate such beam. Our work might be useful in  applications where precise beam control is important such as particle manipulation, filamentation, and micromachining. 
\end{abstract}

% \dates{Compiled \today}

% \ocis{(050.1940) Diffraction; (070.7345) Wave propagation; (260.2110) Electromagnetic optics; (350.7420) Waves.}

%% To be edited by editor
% \doi{\url{http://dx.doi.org/10.1364/XX.XX.XXXXXX}}

\section{Introduction}

The area of curved and accelerating beams has attracted a lot of attention over the last decade. The field initiated with the prediction and experimental observation of Airy beams~\cite{sivil-ol2007,sivil-prl2007}. An Airy beam is characterized by a curved parabolic trajectory as well as diffraction-free and self-healing dynamics. Subsequently, optical waves with engineered arbitrary convex trajectories were generated~\cite{green-prl2011,chrem-ol2011,froeh-oe2011}. Using a different approach, Bessel-like beams that can even bend along non-convex classes of trajectories were observed~\cite{chrem-ol2012-bessel,zhao-ol2013}. The research in this area is triggered by a variety of applications ranging from filamentation~\cite{polyn-science2009,polyn-prl2009} and electric discharge generation~\cite{cleri-sa2015} to particle manipulation~\cite{baumg-np2008,zhang-ol2011,zheng-ao2011,schle-nc2014,zhao-sr2015}, microscopy and imaging~\cite{jia-np2014,vette-nm2014}, and micromachining~\cite{jia-np2014,vette-nm2014}. Interestingly, accelerating waves with radial symmetry were shown to have abrupt focusing properties~\cite{efrem-ol2010}. Such abruptly autofocusing waves have been utilized in particle manipulation~\cite{zhang-ol2011}, creating ablation spots in materials~\cite{papaz-ol2011}, and filamentation~\cite{panag-nc2013}. 

Optical waves with curved trajectories have also been found in the non-paraxial regime with their main advantage being that the trajectory of the beam can bend at large angles~\cite{froeh-oe2011}. The particular case of a circular trajectory is described in terms of Bessel functions and exhibits diffraction-free dynamics~\cite{kamin-prl2012,zhang-ol2012,courv-ol2012}. Additional classes of exact solutions of the Mathieu and Weber type have also been observed~\cite{zhang-prl2012,aleah-prl2012,bandr-njp2013}. Closed-form expressions for the phase profile of different classes of accelerating beams were found  in~\cite{penci-ol2015}. In addition, in the non-paraxial regime the focusing properties of abruptly autofocusing waves can be further enhanced~\cite{penci-ol2016}.

For beams that propagate along straight trajectories, it has been shown that it is possible to control their longitudinal shape by using an equal frequency Bessel beam decomposition~\cite{zambo-oe2004}. Furthermore, Bessel beams with tunable axial profile have been proposed and observed~\cite{cizma-oe2009}. One of the main accomplishments in the area of curved beams is the ability to fully control the trajectory of an accelerating beam. However, there are additional degrees of freedom such as the beam width and the maximum amplitude along the trajectory that up to now remain almost unexplored. In this respect, in~\cite{hu-ol2013-am} an accelerating beam with a periodic amplitude profile was generated by using a periodic modulation to the spectrum. Furthermore, amplitude corrections have been applied in Airy beams to compensate for losses during propagation~\cite{preci-ol2014}. Such a technique has been utilized to improve the imaging capabilities of Airy beam light-sheet microscopy~\cite{nylk-sa2018}. However, no generic method has been proposed that allows full control of the maximum amplitude along the trajectory almost at will.

The purpose of this work is to show that it is possible to independently control both the trajectory and the maximum amplitude along the trajectory of an accelerating wave in the non-paraxial regime. This is achieved by utilizing both the amplitude and the phase degrees of freedom of the initial waveform. In addition, we show that the beam width solely depends on the local curvature of the trajectory. Thus, we can generate beams with designed amplitude and beam width. Our asymptotic calculations are in excellent agreement with direct numerical simulations. The results of this work might be utilized in a variety of applications where precise control of the properties of the beam is important, such as particle manipulation, micromachining, filamentation, and electric discharge generation. 

\section{Amplitude and trajectory/beam-width engineering}

The dynamics of a monochromatic optical light sheet that is polarized along the $\hat{\bm y}$-direction $\bm E = \hat{\bm y}\psi(x,z)e^{-i\omega t}$ and propagates in a linear homogeneous and isotropic dielectric medium is given by the Rayleigh-Sommerfeld integral
\begin{equation}
\psi(x,z) = 2\int_{-\infty}^\infty \psi_0(\xi)
\frac{\partial G(x,z;\xi)}{\partial z}\dif\xi. 
\label{eq:green}
\end{equation}
In Eq.~(\ref{eq:green}) $\psi_0(x)=A(x)e^{i\phi(x)}$ is the optical wave on the input plane, $x$ is the transverse and $z$ is the longitudinal coordinate, $k=n\omega/c=2\pi/\lambda$, $G(x,z;\xi)= -(i/4)H_0^{(1)}(kR)$, $H_0^{(1)}$ is a Hankel function, and $R=((x-\xi)^2+z^2)^{1/2}$. Applying first and second order stationarity of the phase to the Rayleigh-Sommerfeld integral~\cite{penci-ol2015}, we derive the ray equation $x=\xi+R\phi'(\xi)/k$ and the prediction for the beam trajectory. On the other hand, it is possible to solve the inverse problem of finding the phase that is required in order to generate a beam with a convex but otherwise arbitrary trajectory $x_c=f(z_c)$. By noting that a straight line that is tangent to the trajectory is a ray equation, we obtain the following relation for the derivative of the phase~\cite{froeh-oe2011}
\begin{equation}
\phi'(\xi)=
k(\dif f/\dif z_c)/(1+(\dif f/\dif z_c)^2)^{1/2}
\label{eq:tangent}
\end{equation}
where $\xi = f(z_c)-z_cf'(z_c)$. 
The above equation can be analytically integrated for different classes of convex trajectories such as circular, elliptic, power-law, and exponential~\cite{penci-ol2015}. We are going to utilize these closed-form expressions below.

We would like to obtain an asymptotic expression for the dynamics of the optical wave close to the caustic. In order to do so, we first expand the phase of the integrand in Eq.~(\ref{eq:green}) in a Taylor series close to the caustic as $\xi=\xi_c+\delta\xi$, $x=x_c+\delta x$, $z=z_c$ and keep terms up to cubic order. We assume that the amplitude of the incident beam $A(\xi)$ is relatively slowly varying so that the main contribution to the integral comes from phase stationarity. Finally we directly integrate with respect to $\delta\xi$. After some calculations we obtain
\begin{equation}
\psi = 
2A(\xi)
(\pi^4 R_c^3\kappa^2/\lambda)^{1/6}
e^{i\Xi}
\Ai(-s(z_c/R_c)[2k^2\kappa(z_c)]^{1/3}\delta x)
\label{eq:psi:asy}
\end{equation}
where $s=\sgn(z_c')$, $\kappa$ is the curvature of the trajectory
\[
\kappa(z_c) = 
|\dif^2f(z_z)/\dif z_c^2|/[1+(\dif f(z_c)/\dif z_c)^2]^{3/2}
=z_c^2/[R_c^3|z_c'(\xi)|],
\]
$\Xi = kR_c+\phi+k(x_c-\xi) \delta x/R_c+kz_c^2(\delta x)^2/(2R_c^3)-\pi/4$, $R_c=R(x_c,z_c,\xi)$ and for simplicity we have replaced $\xi_c$ with $\xi$. We see that, independent of the type of the trajectory, close to the caustic the optical wave takes the form of an Airy function. In addition, the maximum amplitude in the transverse plane [$U(z_c(\xi))=\{\max|\psi|: x\in\mathbb{R}\}$] is a function that depends both on the amplitude on the input plane and the beam trajectory. Thus, we can determine the maximum amplitude along the trajectory by appropriately selecting the amplitude on the input plane as
\begin{equation}
A(\xi)
=
F(\xi)(U(\xi)/2.3)
(\lambda R_c^3\kappa^2(z_c)/R_c^3)^{1/6},
\label{eq:ampl}
\end{equation}
where $F(\xi)$ is a truncation factor that is zero in the domain where $\phi(\xi)$ [and thus $\kappa(z_c(\xi))$] is not defined and smoothly goes to unity outside this domain.

In the paraxial regime a beam is considered to be diffraction-free if its amplitude remains invariant in the transverse plane. The main spot of a diffraction-free beam can be extremely narrow without being subjected to diffractive spreading. In the non-paraxial regime accelerating beams can bend at large angles and thus the notion of what is diffraction-free might be accordingly modified. Specifically, we can define a beam to be diffraction-free if its amplitude profile remains invariant in the plane that is normal to the trajectory. Since the most important feature of a diffraction-free beam is its main lobe, such a calculation can even be local (valid close to the beam trajectory). We repeat the same type of asymptotic calculations in a coordinate system that follows the beam trajectory $(x_c,z_c)$ where the first coordinate is tangent and the second ($l$) is normal to the trajectory. We find that close to the caustic 
\begin{equation}
\psi = 
2A(\xi)
(\pi^4 R_c^3\kappa^2/\lambda)^{1/6}
e^{i\Xi}
\Ai\left(-s
[2k^2\kappa(z_c)]^{1/3}
\delta l\right).
\label{eq:psi:asy2}
\end{equation}
We come to the elegant result that the beam width depends solely on the curvature of the trajectory
$W(\xi) = 1/[2k^2\kappa(z_c)]^{1/3}$.
As the curvature increases the beam width becomes smaller. Since the only trajectory with constant curvature is the circle, we conclude that only circular trajectories have constant beam widths~\cite{kamin-prl2012,zhang-ol2012,courv-ol2012}. 
Importantly, by engineering the curvature of the trajectory we can generate beams with designed beam widths along the trajectory.

\begin{figure}
\centerline{
\includegraphics[width=\columnwidth]{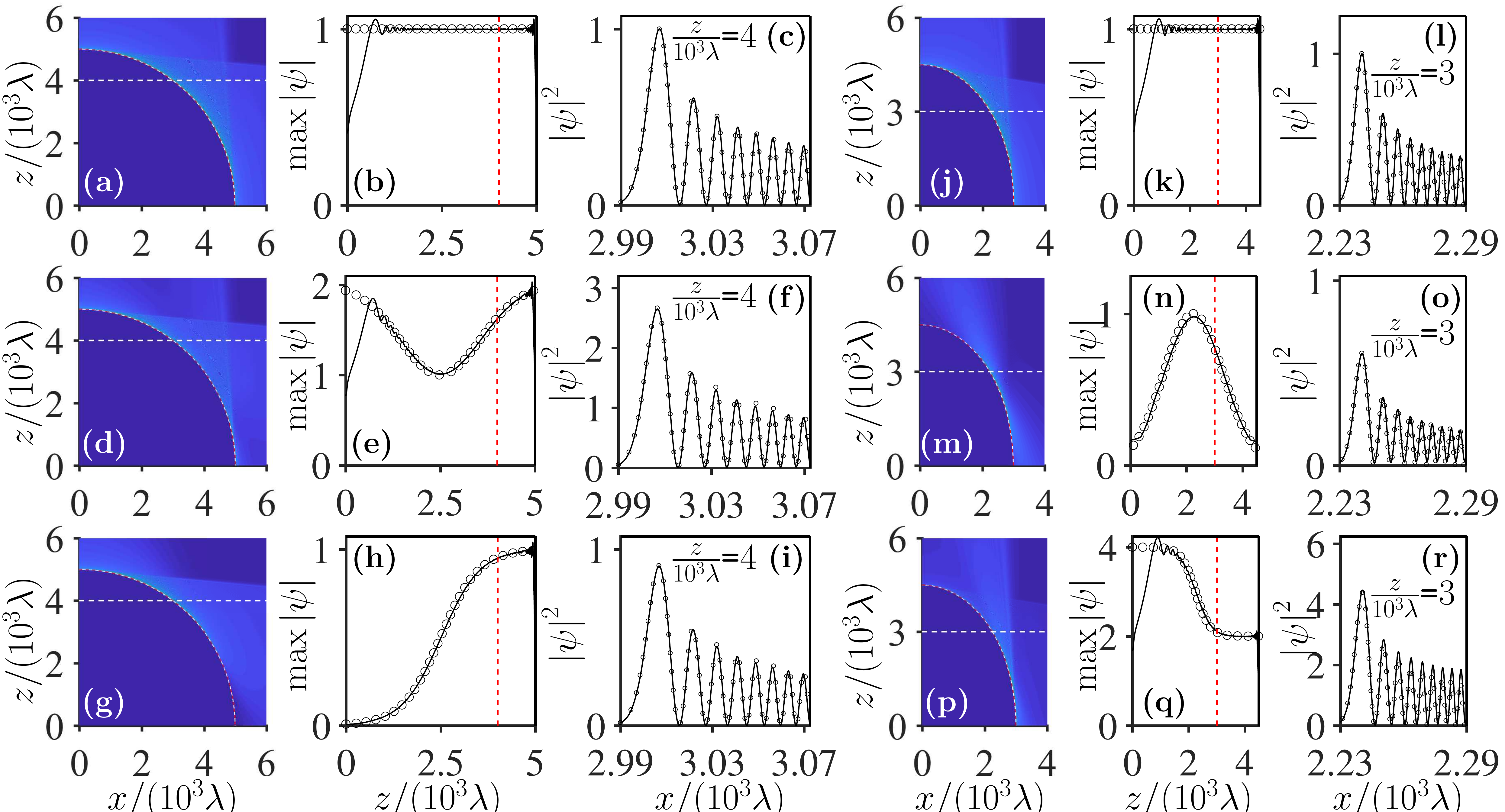}
}
\caption{Accelerating beams with circular (first three columns) and elliptic (last three columns) trajectories. In the two subfigures we depict in the first column the amplitude dynamics and the theoretical prediction for the trajectory (white-red dashed curve), in the second column the maximum amplitude as a function of the propagation distance, and in the third column the intensity profile at the cross section shown in the first column (white dashed line). In the second and third columns the theoretical predictions are shown with circles. For the circular trajectory $x_c=(R_0^2-z_c^2)^{1/2}$, $R_0=5000\lambda$, and the engineered amplitude profile in the three rows is 
$U=1$, $U=[2-\exp(-(z-R_0/2)^2/(1500\lambda)^2)]$, $U=[0.5+0.5\tanh(0.001(z-R_0/2)/\lambda)]$, respectively. 
For the elliptic trajectory $x_c=f(z_c)=(R_0^2-(z_c/\alpha)^2)^{1/2}$, $R_0=3000\lambda$, $\alpha=3/2$ and the amplitude profile in the three rows is 
$U=1$, $U=\exp(-(z-3R_0/4)^2/(1500\lambda)^2)$, $U=[3-2\tanh(0.002(z-3R_0/4)/\lambda)]$, respectively. In all the figures $F(\xi) = \sig[(\xi-R_0)/(5\lambda)]$.
}
\label{fig:1}
\end{figure}

\section{Numerical results}
In all our simulations below the truncation $F(\xi)$ is defined in connection to the sigmoid function 
\[
\operatorname{sig}(x) = \max(\tanh(x),0)
\]
to continuously set the amplitude to zero outside the range where the phase and the amplitude on the input plane are well defined. For arbitrary functions of the convex trajectory $x_c=f(z_c)$ and the maximum amplitude $U(z_c)$ the amplitude and phase on the input plane is, in general, derived numerically by integrating Eqs.~(\ref{eq:tangent}) and~(\ref{eq:ampl}). For the particular examples presented in this paper we are going to utilize classes of trajectories that lead to closed-form expressions~\cite{penci-ol2015}. In Figs.~\ref{fig:1}(a)-\ref{fig:1}(i) we depict typical results of a non-paraxial beam that follows a circular trajectory. 
We select three different cases for the amplitude. In the first row the functional form for the maximum amplitude is set to unity. In Fig~\ref{fig:1}(b) we see that at the first stages of propagation the numerically computed maximum amplitude is small, but rapidly increases with $z$ and after some distance it reaches the desired constant value. After this initial stage and up to $z=R_0$ the comparison between the theoretical and the numerical results is excellent. However, some discrepancies still happen very close to $z=R_0$. The reason for this is that the rays that form the caustic as $z_c$ approaches $R_0$ are emitting from $\xi$ that goes to infinity. Also in Fig.~\ref{fig:1}(c) an intensity cross-section is shown and compared with the analytic expression. It is actually remarkable that Eq.~(\ref{eq:psi:asy}) does not only predict the behavior of the wave close to the caustic, but also accurately describes several lobes of the amplitude profile. In the second and third rows we have made different selections for the maximum amplitude along the trajectory. Again theory and simulations show very good agreement.

\begin{figure}
\centerline{\includegraphics[width=\columnwidth]{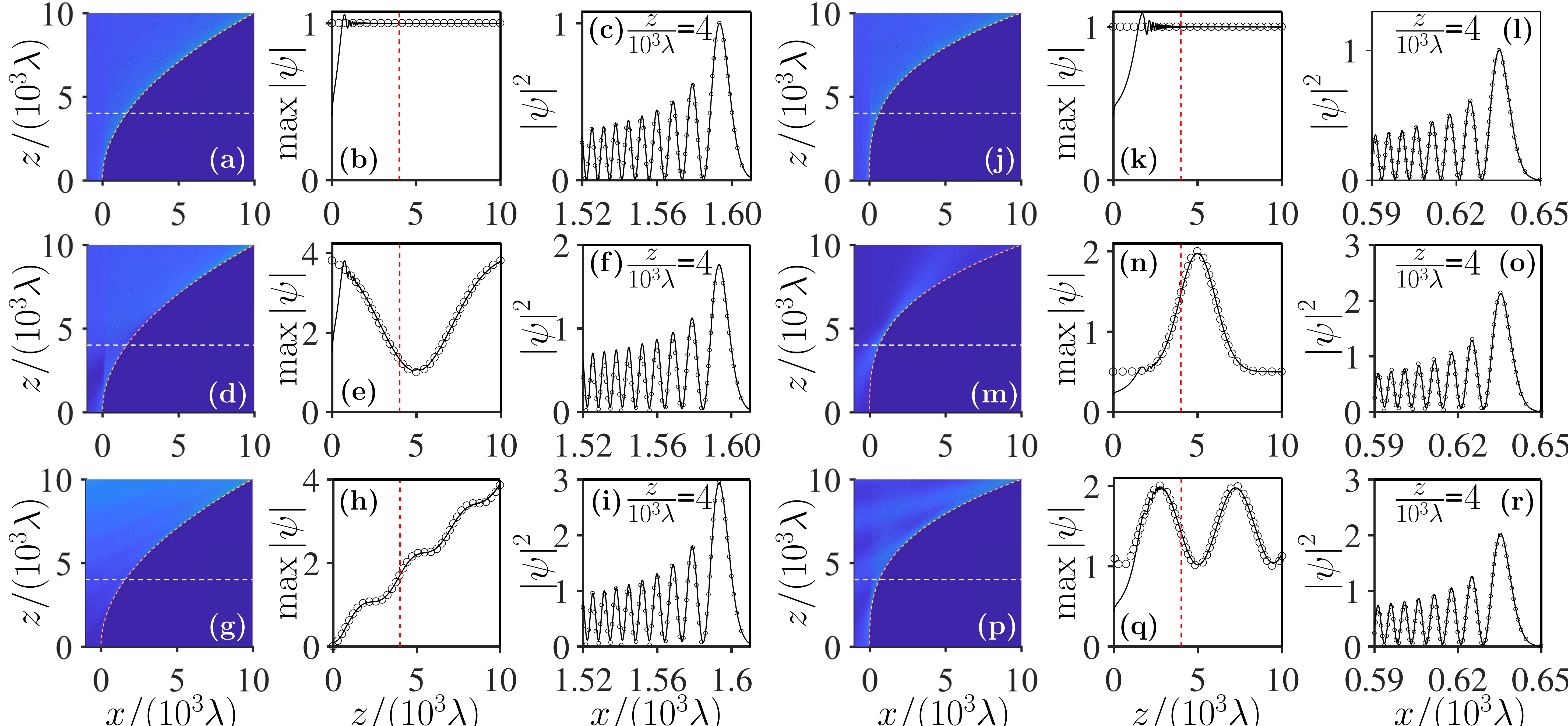}}
\caption{Same as in Fig.~\ref{fig:1}. In the left subfigure we show accelerating beams with a parabolic trajectory $x_c=f(z_c)=\alpha z_c^2$ and $\alpha=10^{-4}/\lambda$. The engineered amplitude profile in the three rows is 
$U=1$, 
$U=[4-3\exp(-(z-5000\lambda)^2/(3000\lambda)^2)]$,
$U=[0.75z/\lambda+750\sin^2(0.001z/\lambda)]/2000$.
In the right subfigure we show accelerating beams with with a cubic trajectory $x_c=f(z_c)=\alpha z_c^3$ and $\alpha=10^{-8}/\lambda^2$. The engineered amplitude profile in the three rows is 
$U=1$, 
$U=[0.5+1.5\exp(-(z-5000\lambda)^2/(1500\lambda)^2)]$,
$U=[1+\sin^2(0.0007(z-5000\lambda))]$,
respectively. In all the figures $F(\xi) = \sig(-\xi/(5\lambda))$.
}
\label{fig:2}
\end{figure}
In addition, we have tested different functional forms for the beam trajectory and the maximum amplitude along the trajectory. In all these cases the comparison between theory and numerics is very good. The only restriction is that the maximum amplitude along the trajectory should be relatively slowly varying. Specifically, in Fig.~\ref{fig:1}(j)-\ref{fig:1}(r) we selected an elliptic nonparaxial trajectory. The results here are comparable to the circular case. Finally, in Fig.~\ref{fig:2} we use power-law trajectories (parabolic and cubic). Since in these cases the trajectories do not bend at a $90^\circ$ angle, the required aperture is finite and thus the comparison between theory and numerics remains very good up to the propagation distance of the simulation. In all these classes of trajectories we have selected to test different functional forms for the maximum amplitude including sinusoidal, Gaussian on top of a pedestal, sigmoid etc. However, a particularly useful case in terms of applications is that of constant maximum amplitude along the trajectory which is shown in the top row of all our Figures. For example, in particle manipulation, it is important to be able to transport particles along a particular trajectory. The particles are trapped in the transverse plane due to intensity gradient forces and get transported due to radiation pressure. Intensity modulations along the beam trajectory are not desirable and might interrupt this process. Note that the third and sixth row in both figures we see the excellent agreement between the theoretical prediction and the numerical results for the amplitude profile not only as it concerns the first lobe (the beam-width) but also for several additional lobes.

\section{Conclusions}
In conclusion, we have shown that it is possible to generate non-paraxial optical waves with engineered amplitude and trajectory/beam width. Our method relies on modulating both the amplitude and the phase of the incident wave. 
Our numerical results are in excellent agreement with the theoretical predictions. 
This work might be useful in different applications where precise control of the beam parameters is important including particle manipulation, filament and electric discharge generation, and micromachining. 

Note that the experimental realization of the proposed optical waves require the encoding of both amplitude and phase information on the input plane. An experimentally simpler approach is to encode the complex waveform either by using binary computer generated holograms~\cite{lee-ao1979,lee-po1978, libst-prl2014} or by using a spatially modulated phase patterns~\cite{davis-ao1999,papaz-ol2011}.

\section*{Funding} 
Supported by the Special Account for Research of the University of Crete and by the Greek State Scholarships Foundation (IKY).

\newcommand{\noopsort[1]}{} \newcommand{\singleletter}[1]{#1}


\begin{thebibliography}{10}
\newcommand{\enquote}[1]{``#1''}

\bibitem{sivil-ol2007}
G.~A. Siviloglou and D.~N. Christodoulides, \enquote{Accelerating finite energy
  {A}iry beams,} Opt. Lett. \textbf{32}, 979--981 (2007).

\bibitem{sivil-prl2007}
G.~A. Siviloglou, J.~Broky, A.~Dogariu, and D.~N. Christodoulides,
  \enquote{Observation of accelerating {A}iry beams,} Phys. Rev. Lett.
  \textbf{99}, 213901 (2007).

\bibitem{green-prl2011}
E.~Greenfield, M.~Segev, W.~Walasik, and O.~Raz, \enquote{Accelerating light
  beams along arbitrary convex trajectories,} Phys. Rev. Lett. \textbf{106},
  213902 (2011).

\bibitem{chrem-ol2011}
I.~Chremmos, N.~K. Efremidis, and D.~N. Christodoulides,
  \enquote{Pre-engineered abruptly autofocusing beams,} Opt. Lett. \textbf{36},
  1890--1892 (2011).

\bibitem{froeh-oe2011}
L.~Froehly, F.~Courvoisier, A.~Mathis, M.~Jacquot, L.~Furfaro, R.~Giust, P.~A.
  Lacourt, and J.~M. Dudley, \enquote{Arbitrary accelerating micron-scale
  caustic beams in two and three dimensions,} Opt. Express \textbf{19},
  16455--16465 (2011).

\bibitem{chrem-ol2012-bessel}
I.~D. Chremmos, Z.~Chen, D.~N. Christodoulides, and N.~K. Efremidis,
  \enquote{{B}essel-like optical beams with arbitrary trajectories,} Opt. Lett.
  \textbf{37}, 5003--5005 (2012).

\bibitem{zhao-ol2013}
J.~Zhao, P.~Zhang, D.~Deng, J.~Liu, Y.~Gao, I.~D. Chremmos, N.~K. Efremidis,
  D.~N. Christodoulides, and Z.~Chen, \enquote{Observation of self-accelerating
  {B}essel-like optical beams along arbitrary trajectories,} Opt. Lett.
  \textbf{38}, 498--500 (2013).

\bibitem{polyn-science2009}
P.~Polynkin, M.~Kolesik, J.~V. Moloney, G.~A. Siviloglou, and D.~N.
  Christodoulides, \enquote{Curved plasma channel generation using ultraintense
  {A}iry beams,} Science \textbf{324}, 229--232 (2009).

\bibitem{polyn-prl2009}
P.~Polynkin, M.~Kolesik, and J.~Moloney, \enquote{Filamentation of femtosecond
  laser {A}iry beams in water,} Phys. Rev. Lett. \textbf{103}, 123902 (2009).

\bibitem{cleri-sa2015}
M.~Clerici, Y.~Hu, P.~Lassonde, C.~Mili{\'a}n, A.~Couairon, D.~N.
  Christodoulides, Z.~Chen, L.~Razzari, F.~Vidal, F.~L{\'e}gar{\'e}, D.~Faccio,
  and R.~Morandotti, \enquote{Laser-assisted guiding of electric discharges
  around objects,} Sci. Adv. \textbf{1}, e1400111 (2015).

\bibitem{baumg-np2008}
J.~Baumgartl, M.~Mazilu, and K.~Dholakia, \enquote{Optically mediated particle
  clearing using {A}iry wavepackets,} Nat. Photon. \textbf{2}, 675--678 (2008).

\bibitem{zhang-ol2011}
P.~Zhang, J.~Prakash, Z.~Zhang, M.~S. Mills, N.~K. Efremidis, D.~N.
  Christodoulides, and Z.~Chen, \enquote{Trapping and guiding microparticles
  with morphing autofocusing {A}iry beams,} Opt. Lett. \textbf{36}, 2883--2885
  (2011).

\bibitem{zheng-ao2011}
Z.~Zheng, B.-F. Zhang, H.~Chen, J.~Ding, and H.-T. Wang, \enquote{Optical
  trapping with focused {A}iry beams,} Appl. Opt. \textbf{50}, 43--49 (2011).

\bibitem{schle-nc2014}
R.~Schley, I.~Kaminer, E.~Greenfield, R.~Bekenstein, Y.~Lumer, and M.~Segev,
  \enquote{Loss-proof self-accelerating beams and their use in non-paraxial
  manipulation of particles' trajectories,} Nat. Commun. \textbf{5}, 5189
  (2014).

\bibitem{zhao-sr2015}
J.~Zhao, I.~D. Chremmos, D.~Song, D.~N. Christodoulides, N.~K. Efremidis, and
  Z.~Chen, \enquote{Curved singular beams for three-dimensional particle
  manipulation,} Sci. Rep. \textbf{5}, 12086 (2015).

\bibitem{jia-np2014}
S.~Jia, J.~C. Vaughan, and X.~Zhuang, \enquote{Isotropic three-dimensional
  super-resolution imaging with a self-bending point spread function,} Nat.
  Photon. \textbf{8}, 302--306 (2014).

\bibitem{vette-nm2014}
T.~Vettenburg, H.~I.~C. Dalgarno, J.~Nylk, C.~Coll-Llad\'o, D.~E.~K. Ferrier,
  T.~Cizmar, F.~J. Gunn-Moore, and K.~Dholakia, \enquote{Light-sheet microscopy
  using an {A}iry beam,} Nat. Methods \textbf{11}, 541--544 (2014).

\bibitem{efrem-ol2010}
N.~K. Efremidis and D.~N. Christodoulides, \enquote{Abruptly autofocusing
  waves,} Opt. Lett. \textbf{35}, 4045--4047 (2010).

\bibitem{papaz-ol2011}
D.~G. Papazoglou, N.~K. Efremidis, D.~N. Christodoulides, and S.~Tzortzakis,
  \enquote{Observation of abruptly autofocusing waves,} Opt. Lett. \textbf{36},
  1842--1844 (2011).

\bibitem{panag-nc2013}
P.~Panagiotopoulos, D.~Papazoglou, A.~Couairon, and S.~Tzortzakis,
  \enquote{Sharply autofocused ring-{A}iry beams transforming into non-linear
  intense light bullets,} Nat. Commun. \textbf{4}, 2622 (2013).

\bibitem{kamin-prl2012}
I.~Kaminer, R.~Bekenstein, J.~Nemirovsky, and M.~Segev, \enquote{Nondiffracting
  accelerating wave packets of {M}axwell's equations,} Phys. Rev. Lett.
  \textbf{108}, 163901 (2012).

\bibitem{zhang-ol2012}
P.~Zhang, Y.~Hu, D.~Cannan, A.~Salandrino, T.~Li, R.~Morandotti, X.~Zhang, and
  Z.~Chen, \enquote{Generation of linear and nonlinear nonparaxial accelerating
  beams,} Opt. Lett. \textbf{37}, 2820--2822 (2012).

\bibitem{courv-ol2012}
F.~Courvoisier, A.~Mathis, L.~Froehly, R.~Giust, L.~Furfaro, P.~A. Lacourt,
  M.~Jacquot, and J.~M. Dudley, \enquote{Sending femtosecond pulses in circles:
  highly nonparaxial accelerating beams,} Opt. Lett. \textbf{37}, 1736--1738
  (2012).

\bibitem{zhang-prl2012}
P.~Zhang, Y.~Hu, T.~Li, D.~Cannan, X.~Yin, R.~Morandotti, Z.~Chen, and
  X.~Zhang, \enquote{Nonparaxial {M}athieu and {W}eber accelerating beams,}
  Phys. Rev. Lett. \textbf{109}, 193901 (2012).

\bibitem{aleah-prl2012}
P.~Aleahmad, M.-A. Miri, M.~S. Mills, I.~Kaminer, M.~Segev, and D.~N.
  Christodoulides, \enquote{Fully vectorial accelerating diffraction-free
  {H}elmholtz beams,} Phys. Rev. Lett. \textbf{109}, 203902 (2012).

\bibitem{bandr-njp2013}
M.~A. Bandres and B.~M. Rodr\'iguez-Lara, \enquote{Nondiffracting accelerating
  waves: {W}eber waves and parabolic momentum,} New J. Phys. \textbf{15},
  013054 (2013).

\bibitem{penci-ol2015}
R.-S. Penciu, V.~Paltoglou, and N.~K. Efremidis, \enquote{Closed-form
  expressions for nonparaxial accelerating beams with pre-engineered
  trajectories,} Opt. Lett. \textbf{40}, 1444--1447 (2015).

\bibitem{penci-ol2016}
R.-S. Penciu, K.~G. Makris, and N.~K. Efremidis, \enquote{Nonparaxial abruptly
  autofocusing beams,} Opt. Lett. \textbf{41}, 1042--1045 (2016).

\bibitem{zambo-oe2004}
M.~Zamboni-Rached, \enquote{Stationary optical wave fields with arbitrary
  longitudinal shape by superposing equal frequency bessel beams: Frozen
  waves,} Opt. Express \textbf{12}, 4001--4006 (2004).

\bibitem{cizma-oe2009}
T.~\v{C}i\v{z}m\'{a}r and K.~Dholakia, \enquote{Tunable bessel light modes:
  engineering the axial propagation,} Opt. Express \textbf{17}, 15558--15570
  (2009).

\bibitem{hu-ol2013-am}
Y.~Hu, D.~Bongiovanni, Z.~Chen, and R.~Morandotti, \enquote{Periodic
  self-accelerating beams by combined phase and amplitude modulation in the
  {F}ourier space,} Opt. Lett. \textbf{38}, 3387--3389 (2013).

\bibitem{preci-ol2014}
M.~A. Preciado, K.~Dholakia, and M.~Mazilu, \enquote{Generation of
  attenuation-compensating {A}iry beams,} Opt. Lett. \textbf{39}, 4950--4953
  (2014).

\bibitem{nylk-sa2018}
J.~Nylk, K.~McCluskey, M.~A. Preciado, M.~Mazilu, Z.~Yang, F.~J. Gunn-Moore,
  S.~Aggarwal, J.~A. Tello, D.~E. Ferrier, and K.~Dholakia,
  \enquote{Light-sheet microscopy with attenuation-compensated
  propagation-invariant beams,} Sci. Adv. \textbf{4}, eaar4817 (2018).

\bibitem{lee-ao1979}
W.-H. Lee, \enquote{Binary computer-generated holograms,} Appl. Opt.
  \textbf{18}, 3661--3669 (1979).

\bibitem{lee-po1978}
W.-H. Lee, \enquote{Computer-generated holograms: Techniques and applications,}
  Prog. Opt. \textbf{16}, 119--232 (1978).

\bibitem{libst-prl2014}
A.~Libster-Hershko, I.~Epstein, and A.~Arie, \enquote{Rapidly accelerating
  {M}athieu and {W}eber surface plasmon beams,} Phys. Rev. Lett. \textbf{113},
  123902 (2014).

\bibitem{davis-ao1999}
J.~A. Davis, D.~M. Cottrell, J.~Campos, M.~J. Yzuel, and I.~Moreno,
  \enquote{Encoding amplitude information onto phase-only filters,} Appl. Opt.
  \textbf{38}, 5004--5013 (1999).

\end{thebibliography}
\end{document}